\documentclass[twocolumn]{svjour3}          
\smartqed  
\usepackage{graphicx}
\usepackage{bm}
\usepackage{amsmath}
\usepackage{amssymb}
\usepackage{color}
\begin{document}

\title{Drag acting on an intruder in a three-dimensional granular environment}

\author{Satoshi Takada \and Hisao Hayakawa}

\institute{
	Satoshi Takada 
	\at Earthquake Research Institute, The University of Tokyo, 1--1--1, Yayoi, Bunkyo-ku, Tokyo 113--0032, Japan\\
	Present address:
	Institute of Engineering, Tokyo University of Agriculture and Technology,
	2--24--16, Naka-cho, Koganei, Tokyo 184--8588, Japan
	\email{takada@go.tuat.ac.jp}           
	\and
	Hisao Hayakawa 
	\at
	Yukawa Institute for Theoretical Physics, Kyoto University, Kitashirakawa Oiwakecho, Sakyo-ku, Kyoto 606--8502, Japan\\
	\email{hisao@yukawa.kyoto-u.ac.jp} 
}

\date{Received: date / Accepted: date}

\maketitle
\abstract{%
The drag acting on an intruder in a three-dimensional frictionless dry granular environment is numerically studied.
It is found the followings: (i) There is no yield force for the motion of the intruder without the gravity.
(ii) The drag is proportional to the cross section of the moving intruder.
(iii) If the intruder is larger than surrounding grains, the drag is proportional to the moving speed $V$ of the intruder for dense systems, but it exhibits a crossover from quadratic to linear dependences of the moving speed when the volume fraction of the surrounding grains is much lower than the jamming point.
(iv) There is a plateau regime where the drag is almost independent of $V$ if the size of the intruder is identical to those of the environmental grains and the volume fraction is near the jamming point.}

\noindent
{\bf keywords:} Drag, intruder, jamming
\section{Introduction}

Dense granular materials behave as solid-like if the density is above the jamming point, while they behave as liquid-like if the density is lower than the jamming point \cite{Liu98,Behringer18}.
It is known that the jamming transition of frictionless granular grains is a mixed transition in which the coordination number is discontinuously changed from zero to finite at the jamming point in the hard core limit, while the stress is continuously changed at the jamming point \cite{O'Hern02}.
It is also known that the continuous transition of the stress satisfies a critical scaling which is similar to the critical phenomena in the second order phase transition \cite{Olsson07,Hatano08,Otsuki09a,Otsuki09b,Tighe10}. 

So far, most of the studies on the jamming discuss homogeneous compressions or uniformly sheared states of grains.
There is another simple setup which differs from the conventional ones to characterize the rheological response of dense granular materials, {\it i.~e.} the investigation of the drag acting on an intruder in a granular environment.
It is easy to control the drag on the intruder, which is an advantage to study this setup.
Indeed, for instance, the sheared dry granular systems are always suffered from shear bands and non-uniformity which prevents the system from idealistic situations as assumed in theoretical studies.

There are many studies to characterize the drag acting on a moving intruder in a  granular environment \cite{Albert99,Albert01,Chehata03,Wassgren03,Bharadwaj06,Geng04,Geng05,Reddy11,Hilton13,Potiguar13,Guillard13,Guillard15,Takehara10,Takehara14,Takada17,Kumar17,Brzinski10}.
Recent experiments in quasi two-dimensional systems \cite{Takehara10,Takehara14} clarified that the resistance force consists of the yield force which is independent of the moving speed and the dynamic term proportional to $V^2$ where $V$ is the moving speed of the intruder \cite{Wassgren03,Bharadwaj06,Takada17}.
They also demonstrated that the drag diverges at the jamming point, though the exponents for divergences of the yield force and the drag coefficient in front of $V^2$ are much smaller than those for sheared systems. 
Takada and Hayakawa numerically studied the drag under the corresponding geometry in Refs.\ \cite{Takehara10,Takehara14} to reproduce the results of Ref.\ \cite{Takehara14} and found that the yield force exists only if there exists the friction between the intruder and the bottom plate \cite{Takada17}.
On the other hand, the dynamical drag term is proportional to the effective cross section of the moving intruder and $V^2$, which can be explained by a simple collision model \cite{Wassgren03,Bharadwaj06,Panaitescu17,Jewel18}.
The drag law in dry two-dimensional granular environments seems to be different from that for the Stokesian drag law in viscous fluids which is proportional to $V$ and the diameter of the intruder.

We little know on the drag acting on an intruder in three-dimensional dry granular environments, though the drag law acting on intruder immersed in a mixture of grains and liquids \cite{Nucci19} or in a mixture of grains and the air flow \cite{Brzinski10} is relatively known.
Contrast to two-dimensional cases, the motion of the intruder is strongly affected by the gravity.
In such a system, the resistance force acting on the intruder consists of the yield force and the dynamical drag term which is proportional to the moving speed $V$ \cite{Hilton13}.
To observe a nonlinear contribution of $V$ on the dynamical drag, we may need a large driving force or a high moving speed \cite{Kumar17}.
On the other hand, an experiment on a moving cylinder in a three-dimensional granular media under the Taylor-Couette configuration suggests that the drag $F_{\rm drag}$ obeys a logarithmic function of the moving speed as $F_{\rm drag}-F_c \sim \ln (V/V_0)$ where $F_c$ and $V_0$ are the characteristic force and speed, respectively \cite{Geng05,Reddy11,Candelier09}.
This logarithmic law is completely different from the conventional drag law which can be described by an algebraic function of $V$.
To clarify the drag law acting on the intruder in three-dimensional dry granular environments, we have to analyze simple and idealistic situations.

The goal of this paper is to clarify the followings;
(i) the origin of the yield force, (ii) the distinction of the linear drag term proportional to $V$ from the quadratic drag term proportional to $V^2$ and (iii) the condition for the appearance of the logarithmic term observed in Ref.\ \cite{Reddy11}.
For this purpose, we adopt an idealistic model in which the grains are frictionless and the effect of gravity is absent.
Namely, we can control the volume fraction (the density) of the environmental grains.
Note that the control of the density of a dry granular system is impossible if the gravity exists.
Even in the absence of the gravity, there are many metastable states depending on the protocol if the mutual friction between grains exists. 
As will be explained in the next section, we control the driving force acting on the intruder and examine some diameters of the intruders in several densities of surrounding grains.

The organization of this paper is as follows.
In the next section, we briefly explain our model and the setup of our simulation.
Section \ref{sec:results} is the main part of this paper, where
we present our numerical results on the drag acting on the intruder for various densities of surrounding grains and diameters of the intruders.
In section 4, we discuss and summarize our results.


\begin{figure}
	\centering
	\includegraphics[width=0.8\linewidth]{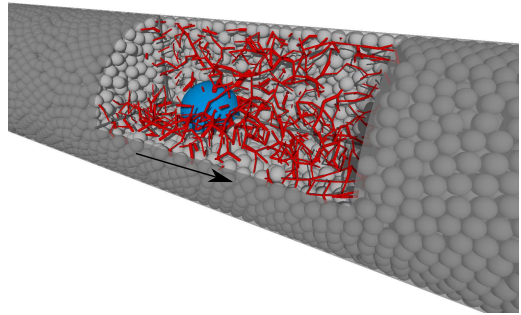}
	\caption{A snapshot of our simulation for $\varphi=0.620$, $D=5d$, and $F_{\rm ex}=5.0\times10^{-2}kd$ by using a surface cut plot of the three-dimensional system.
		The arrow indicates the moving direction of the intruder.
	The random solid (red) lines express the force chains, where their widths are proportional to the magnitude of the interaction in the normal direction.}
	\label{fig:setup}      
\end{figure}

\section{Setup}\label{sec:setup}

In this section, we explain the setup of our molecular dynamics simulation.
The environmental grains consisting of monodisperse frictionless grains (mass $m$ and diameter $d$ for each grain) are confined in a cylinder with the radius $R=7.5d$ and the length $L_x=60d$ for $D\le 5d$, while $R=15d$ and $L_x=120d$ for $D=10d$, where $D$ is the diameter of the intruder.
The intruder is located in the center of the system at the beginning of our simulation and is pulled in the positive $x-$direction (the axial direction) after we start the simulation.
We examine several volume fractions $\varphi$ of the surrounding particles of the systems ranged from $\varphi=0.550$ to $0.645$, where $\varphi$ is defined by $\varphi\equiv \{(\pi/6)d^3N\}/\{\pi R^2L_x - (\pi/6)D^3\}$ with the number of surrounding particles $N$.
The numbers of surrounding grains for various $D/d$ and $\varphi$ are listed in Table \ref{fig:num_particles}.
We assume that the grains are at rest at the beginning of our simulation.
Note that we do not simulate the motion of intruder for low density $\varphi<0.55$ because (i) such low density situations are unreachable for dry granular systems if the gravity exists and (ii) it is hard to reach a steady state in such a low density situation.
We also investigate the drag for $\varphi>\varphi_{\rm J}$ where $\varphi_{\rm J}$ is the jamming fraction ($\varphi_{\rm J}\approx 0.639$).

The equation of motion of $i-$th grain (the intruder or a surrounding grain) located at $\bm{r}_i$ and its acceleration $\ddot{\bm{r}}_i$ is given by
\begin{equation}
	m_i \ddot{\bm{r}}_i = F_{\rm ex}\bm{e}_x\delta_{i,0} +{\sum_{j=0,j\ne i}^{N}}\bm{f}_{ij},
\end{equation}
where the intruder ($i=0$) has the mass $m_0=(D/d)^3m$, each surrounding grain for $i\ge 1$ has the identical mass $m_i=m$, and $\bm{e}_x$ is the unit vector parallel to $x-$direction.
Note that the external force $F_{\rm ex}$ acting only on the intruder is reduced to the drag $F_{\rm drag}$ in steady states. 
The interaction force contains the summation $\sum_{j=0,j\ne i}^{N}$ for all $j$ except for $j=i$ and the explicit form of $\bm{f}_{ij}$ is given by $\bm{f}_{ij}=\Theta(d_{ij}-r_{ij})\{k (d_{ij}-r_{ij})\hat{\bm{r}}_{ij} - \zeta (\bm{v}_{ij}\cdot \hat{\bm{r}}_{ij})\hat{\bm{r}}_{ij}\}$
 with Heaviside's step function $\Theta(x)=1$ for $x\ge 0$ and $\Theta(x)=0$ otherwise.
 Here we have introduced $\bm{r}_{ij}=\bm{r}_i-\bm{r}_j$, $r_{ij}=|\bm{r}_{ij}|$, $\hat{\bm{r}}_{ij}=\bm{r}_{ij}/r_{ij}$, $\bm{v}_{ij}=\bm{v}_i-\bm{v}_j$ with $\bm{v}_i=d\bm{r}_i/dt$, and $d_{ij}=(d_i+d_j)/2$ with $d_0=D$ and $d_i=d$ for $i\ge 1$.
For simplicity, we adopt the linear spring model with the spring constant $k$ instead of the realistic Hertzian contact model for $\bm{f}_{ij}$.
We adopt $\zeta=0.10\sqrt{mk}$ for the viscous constant corresponding to the restitution constant $e=0.8$ in the most of the simulations.
We control the external force $F_{\rm ex}$ and measure a steady value of the velocity of the intruder $V$.
Because the instantaneous moving speed fluctuates, 
we use its time-averaged value for $V$ in the steady state.
We adopt the time step $\Delta t=1.0\times 10^{-3}\sqrt{m/k}$ in most of our simulations.
We have confirmed that the results are unchanged when we change the value of $\Delta t$.

\begin{table}
	\centering
	\caption{A set of parameters (the volume fraction $\varphi$, the diameter of the intruder $D$, and the number of surrounding grains $N$) used in our simulation.}
 	\begin{tabular}{c|c|c||c|c|c}
 	\hline\hline
 	$\varphi$ & $D/d$ & $N$ & $\varphi$ & $D/d$ & $N$ \\ \hline
 	$0.550$ & $5$ & $11,132$ & $0.635$ & $1$ & $12,075$ \\ \hline
 	$0.600$ & $1$ & $12,149$ & $0.635$ & $3$ & $12,478$ \\ \hline
 	$0.600$ & $3$ & $12,554$ & $0.635$ & $5$ & $12,779$\\ \hline
 	$0.600$ & $5$ & $12,858$ & $0.635$ & $10$ & $102,235$ \\ \hline
 	$0.620$ & $1$ & $12,134$ & $0.640$ & $5$ & $12,954$ \\ \hline
 	$0.620$ & $3$ & $12,538$ & $0.645$ & $5$ & $13,055$ \\ \hline
 	$0.620$ & $5$ & $12,842$ &&& \\ 
 	\hline\hline
	\end{tabular}
	\label{fig:num_particles}
\end{table}

\section{Numerical Results}\label{sec:results}

In this section, we present our numerical results. 
First, we show a typical snapshot of our system, where the force chains exist around the intruder (Fig.~\ref{fig:setup}). 
Overall, our results do not have any yield force,
which is consistent with the result of the two-dimensional simulation~\cite{Takada17}.
This suggests that the yield force observed in the previous studies~\cite{Hilton13,Kumar17} is the result of the gravity or the frictional force between the grains and the boundary~\footnote{We have already confirmed that the yield force becomes finite once we have introduced the mutual frictional force between grains under the bumpy boundary condition. Therefore, the mutual friction is more important than the gravity for the yield force.}

\begin{figure}
	\centering
	\includegraphics[width=0.8\linewidth]{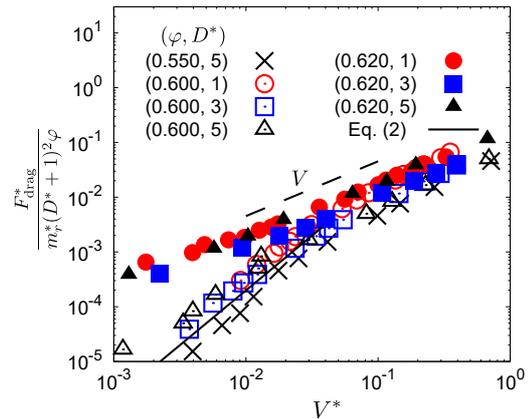}
	\caption{Plots of the drag against the moving speed of the intruder for various  $D/d$ and $\varphi$.
	The dashed line is the guideline which is proportional to $V$, while the solid line expresses Eq.\ \eqref{eq:collision_model}) without any fitting parameter.
	Here, we have introduced the dimensionless speed and drag as $V^*\equiv V/(d\sqrt{k/m})$ and $F_{\rm drag}^*\equiv F_{\rm drag}/(kd)$, respectively, with $m_r^*\equiv m_r/m$ and $D^*\equiv D/d$.
	It is noted that the error bars are smaller than the symbols.}
	\label{fig:F_vs_V_1}      
\end{figure}

Second, we present the drag against $V^*\equiv V/(d\sqrt{k/m})$ for $\varphi=0.55$, $0.60$, and $0.62$ in the cases of $D/d=1, 3$ and 5 (Fig.~\ref{fig:F_vs_V_1}) with $10$ ensembles under different initial conditions.
The results strongly depend on $\varphi$, where the drag is proportional to $V$ in the whole range of $V$ for $\varphi=0.62$ while the drag exhibits the crossover from the region proportional to $V^2$ to that proportional to $V$ for $\varphi=0.55$ and $0.60$.
It is remarkable that the drag can be scaled by the effective cross section $\pi(D+d)^2$ for each $\varphi$.
Note that error bars for all data points are smaller than the sizes of symbols.

\begin{figure}
	\centering
	\includegraphics[width=0.8\linewidth]{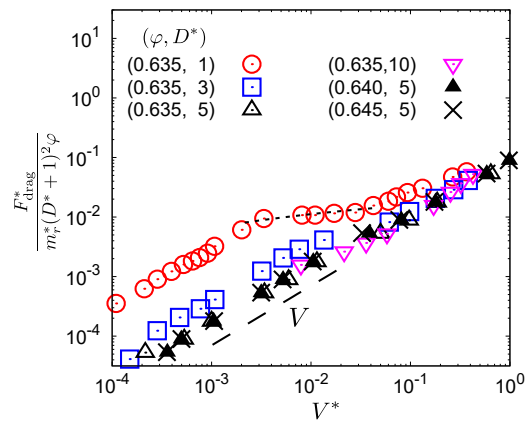}
	\caption{Plots of the drag against the moving speed of the intruder for various $D/d$ and $\varphi=0.635$, $0.640$, and $0.645$.
	The dashed line and the dotted line are, respectively, the guidelines which are proportional to $V$ and the logarithmic function $F_{\rm drag}^*=a+b \ln (cV^*)$ with $a=0.021$, $b=0.0020$, and $c=0.80$.
	It is noted that the error bars are smaller than the symbols.}
	\label{fig:F_vs_V_2}     
\end{figure}

The quadratic dependence for low density can be understood by a simple collision model \cite{Wassgren03}.
Because the momentum change of the $x$-direction for each collision is $\Delta p_x = (1+e)m_rV \cos^2\theta$ with the reduced mass $m_r\equiv Mm/(M+m)$ and the polar angle $\theta$ between the connecting two centers of the masses for contacting grains and the pulling direction ($x$--direction), and the volume of the collision cylinder per unit time is $\Omega_{\rm coll} =(\pi/4)(D+d)^2V$, the force acting on the intruder is given by
\begin{align}
	F_{\rm drag}
	&= \int_0^\pi \sin^2\theta d\theta d\phi \Delta p_x \Omega_{\rm coll}\nonumber\\
	&= (1+e)m_r\frac{(D+d)^2}{d^3} \varphi V^2. \label{eq:collision_model}
\end{align}
Equation \eqref{eq:collision_model} recovers the results of our simulation quantitatively in the quadratic regime for $\varphi\lesssim0.600$ as shown in Fig.\ \ref{fig:F_vs_V_1} without any fitting parameters.
It should be noted that some previous experiments also reported that the drag is proportional to the cross section \cite{Panaitescu17,Jewel18}.

As can be seen in Fig.\ \ref{fig:F_vs_V_2}, there is no regime in which the drag is proportional to $V^2$ for $\varphi\gtrsim0.635$.
It is interesting that the drag strongly depends on $D/d$ for $\varphi\gtrsim0.635$, though the drag is almost independent of $\varphi$ in this dense regime as long as we have checked.
Indeed, there is a plateau for $D/d=1$ in the middle of $V$ or $F^*/\{m_r^*(D^*+1)\varphi\}\simeq 10^{-2}$, where $m_r^*\equiv m_r/m$.
Although the plateau for $D/d\ge 3$ is not clearly visible, the branches for low $V$ and high $V$ are distinguishable, at least, for $D/d\le 3$.
We also note that error bars in all data points in Fig.\ \ref{fig:F_vs_V_2} are smaller than the sizes of symbols.

\begin{figure}
	\centering
	\includegraphics[width=0.8\linewidth]{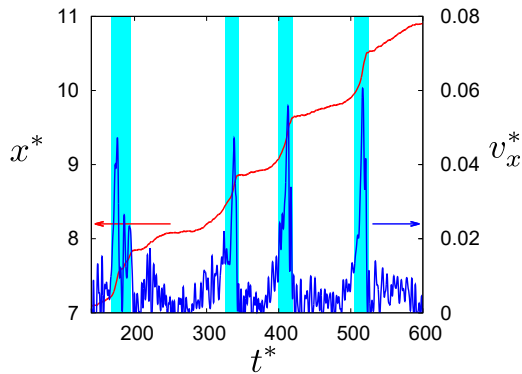}
	\caption{Time evolution of the position and the velocity of the intruder for
	$D/d=1$, $\varphi=0.635$, and $F_{\rm drag}=0.036kd$ with $x^*\equiv x/d$, $v_x^*\equiv v_x/(d\sqrt{k/m})$, and $t^*\equiv t/\sqrt{m/k}$.
	The shaded regions stand for active regions, in which the intruder hops from trapped configurations and the configurations of surrounding grains are rearranged.}
	\label{fig:activation}       
\end{figure}

In the plateau regime, the motion of the intruder is intermittent (Fig.~\ref{fig:activation} for the time evolutions of the position and the velocity of the intruder for $D/d=1$), 
where the motion of the intruder is similar to a stick-slip motion.
This suggests that the motion in this regime is governed by fluctuating hopping processes.
Indeed, the value of the plateau of the drag is almost equal to
$\langle F_c\rangle:=\langle \sum_{\rm contact} k \delta_{ij} \cos\theta \rangle$,
where $\langle \cdots \rangle$ and $\delta_{ij}$ are the ensemble and the contact points average and the overlap $\delta_{ij}\equiv d_{ij}- r_{ij}$, respectively.
We also stress that the drag in the plateau can be fitted by a logarithmic function of $V$ as $F_{\rm drag}^*\equiv F_{\rm drag}/(kd)=a+b\ln (cV^*)$ with $a=0.021$, $b=0.0020$, and $c=0.80$, which is consistent with the observations in Refs.~\cite{Geng05,Reddy11,Candelier09} (Fig.\ \ref{fig:F_vs_V_2}).

\begin{figure}
	\centering
	\includegraphics[width=\linewidth]{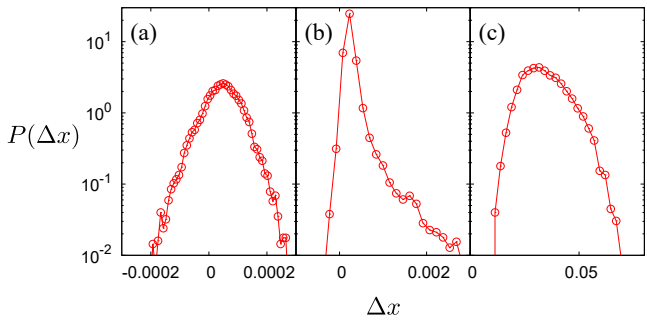}
	\caption{The hopping distribution of the intruder for (a) $F_{\rm ex}=2\times 10^{-3}\ kd$, (b) $2\times 10^{-2}\ kd$, and (c) $0.5\ kd$ for $\varphi=0.635$, where the displacement $\Delta x$ is measured with the time interval $\Delta t=0.1\sqrt{m/k}$.}
	\label{fig:displacement}       
\end{figure}

\begin{figure}
	\centering
	\includegraphics[width=0.8\linewidth]{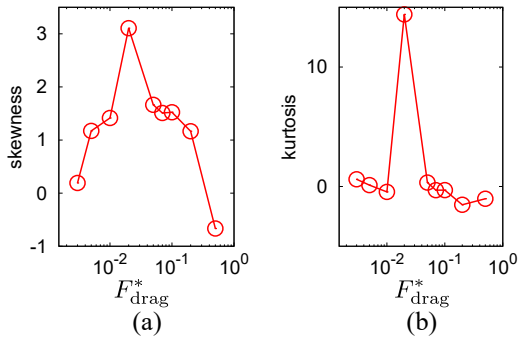}
	\caption{(a) The skewness and (b) the kurtosis of the distributions of the intruder for various $F_{\rm drag}$ for $\varphi=0.635$.}
	\label{fig:skewness_kurtosis}     
\end{figure}

Figure \ref{fig:displacement} exhibits the displacement distributions in the case of $D/d=1$ for several $F_{\rm ex}$.
In the plateau regime, $P(\Delta x)$ has a sharp peak with large skewness and kurtosis, which completely differs from the Gaussian (Fig.\ \ref{fig:displacement}(b)).
The displacement distribution $P(\Delta x)$ obeys a Gaussian for low $F_{\rm ex}$.
Although there exists finite skewness in $P(\Delta x)$, the kurtosis seems to be small for large $F_{\rm ex}$ (Fig.\ \ref{fig:displacement}(c)).
Note that the skewness and kurtosis are
defined by $\mu_3\equiv \kappa_3/\kappa_2^{3/2}-3$  and $\mu_4\equiv \kappa_4/\kappa_2^2$ with the $n$-th moment $\kappa_n\equiv \langle (\Delta x-\langle \Delta x\rangle)^n\rangle$ , where $\langle \Delta x\rangle$ is the ensemble average of $\Delta x$.
Therefore, the motions of the intruder for these regimes are almost determined by the force balance, where the fluctuations are not important.
On the other hand, $P(\Delta x)$ in the plateau regime is quite characteristic,
where there exists a singular peak and the skewness and the kurtosis are quite large.
This result also supports that the motion of the intruder is governed by the fluctuations (see Fig.\ \ref{fig:skewness_kurtosis}).

Let us summarize the observed drag in our simulation:
For $\varphi\lesssim0.600$ and $V^*\lesssim10^{-1}$, 
\begin{equation}\label{eq:2}
	F_{\rm drag}\propto (D+d)^2 V^2.
\end{equation}
When $D/d=1$ and $10^{-2.5}\le V^*\le 10^{-1.5}$, we observe the logarithmic drag
\begin{equation}\label{eq:3}
	F_{\rm drag}\propto \ln V^*.
\end{equation}
For the high speed regime $V^*\ge 10^{-1}$ or the dense regime with $D/d\gg 1$, we observe the linear drag law
\begin{equation}\label{eq:4}
	F_{\rm drag} \propto (D+d)^2 V.
\end{equation}
Note that the constant drag in Eq.~\eqref{eq:3} can be fitted by a logarithmic function as shown in Fig.\ \ref{fig:F_vs_V_2}, which is consistent with those observed in Refs.~\cite{Geng05,Reddy11,Candelier09}.
We, thus, believe that our simulation gives a proper answer to the original questions listed in the introduction for various force laws observed in previous studies.

\section{Discussion and Conclusion}
\begin{figure}
	\centering
	\includegraphics[width=0.8\linewidth]{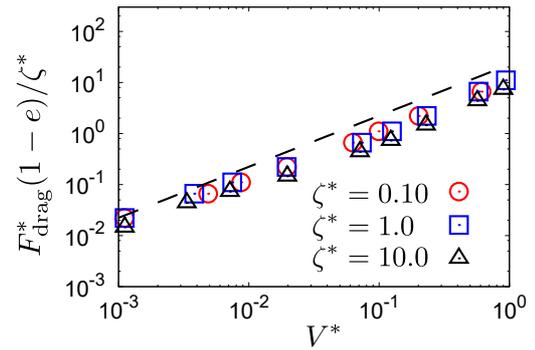}
	\caption{The dissipation rate dependence of the drag for $\zeta^*=0.10$, $1.0$, and $10.0$ with $\zeta^*\equiv \zeta/\sqrt{mk}$.
	Each value corresponds to $e=0.80$, $0.043$, and $0.0048$, respectively.
	The dashed line represents Eq.\ \eqref{eq:linear}.}
	\label{fig:F_drag_V_zeta}       
\end{figure}

It is reasonable that the drag is proportional to $V^2$ in granular systems, because the force is proportional to $({\rm time})^{-2}$ while $d/V$ is only the time scale in the hard core limit.
Therefore, the drag proportional to $V$ appeared in Figs.\ \ref{fig:F_vs_V_1} and \ref{fig:F_vs_V_2} should be scaled by a contact force which contains time scales $\sqrt{m/k}$ and $m/\zeta$.
As long as we have checked, the drag is independent of $k$ or $\Delta t^*$,
while the result strongly depends on $\zeta$.
Therefore, we conclude that the drag proportional to $V$ is dominated by $\zeta$ (see Fig.\ \ref{fig:F_drag_V_zeta}). 
This result also suggests that the drag is proportional to the combination of $V$ and $V_\zeta\equiv \zeta d/\{m(1-e^2)\}$ as
\begin{align}
	F_{\rm drag}&=(1+e)m_r \frac{(D+d)^2}{d^3}\varphi V_\zeta V\nonumber\\
	&=\frac{\zeta}{1-e}\frac{m_r}{m}\frac{(D+d)^2}{d^2}\varphi V, \label{eq:linear}
\end{align} 
which is the origin of the linear regime in Figs.\ \ref{fig:F_vs_V_1} and \ref{fig:F_vs_V_2}.

In summary, we have performed the molecular dynamics simulations for the motion of an intruder in three-dimensional frictionless granular environments to clarify the drag acting on the intruder.
There is no yield force in our setup.
We have found that the drag is proportional to the cross section.
The drag exhibits a crossover from the regime proportional to $V^2$ to that proportional to $V$ for $\varphi\lesssim0.600$, while the drag is always proportional to $V$ for $\varphi=0.620$ and $\varphi\ge0.635$ with $D/d\ge 3$. 
There exists a plateau which is consistent with the logarithmic law observed in Ref.~\cite{Reddy11} for $D/d=1$ and $\varphi=0.635$.
In the plateau regime, the motion of the intruder is intermittent and is governed by the fluctuations. 

To clarify the role of the mutual friction force between contacted grains will be important.
Indeed, sheared frictional granular materials show some exotic phenomena such as the discontinuous shear thickening, the shear jamming as well as the existence of the fragile phase~\cite{Otsuki11,Bi11,Otsuki18}.
Therefore, we expect that the motion of intruder in a frictional granular environment also becomes exotic.
To clarify the origin of the yield force, whether the gravity or the mutual friction, will be an important issue to be solved in the near future.


\begin{acknowledgements}
This work is partially supported by the Grant-in-Aid of MEXT for Scientific Research (Grant No. 16H04025). 
One of the authors (ST) appreciates the warm hospitality of Yukawa Institute for Theoretical Physics at Kyoto University during his stay there supported by the Programs YITP-T-18-03 and YITP-W-18-17.
We would like to dedicate this paper to the memory of Robert P. Behringer who has developed this field.
\end{acknowledgements}

\end{document}